\documentclass[useAMS,usenatbib]{mn2e}

\usepackage{graphics,graphicx}

\title[Multiwavelength Study of {\it Chandra} X-Ray Sources in the Antennae]{Multiwavelength Study of {\it Chandra} X-Ray Sources in the Antennae}

\author[D. M. Clark, et al.]{D. M. Clark$^{1,2}$, S. S. Eikenberry
  $^{2}$, B. R. Brandl$^{3}$, J. C. Wilson$^{4}$,
  J. C. Carson$^{5,6}$, \newauthor C. P. Henderson$^{7}$, T. L. Hayward$^{8}$,
  D. J.  Barry$^{7}$, A. F. Ptak$^{9}$ and E. J. M. Colbert$^{10}$ \\
  $^{1}$Instituto de Astronom\'{i}a, Universidad Nacional Aut\'{o}noma
  de M\'{e}xico, Apdo Postal 877, Ensenada, Baja California,
  M\'{e}xico; \\ dmclark@astrosen.unam.mx \\
  $^{2}$Department of Astronomy, University of Florida, Gainesville, FL 32611 \\
  $^{3}$Leiden University, P.O. Box 9513, 2300 RA Leiden, Netherlands \\
  $^{4}$Department of Astronomy, P.O Box 400325, University of Virginia, Charlottesville, VA 22904 \\
  $^{5}$Max Planck Institute for Astronomy, K\"{o}nigstuhl 17, D-69117 Heidelberg, Germany \\
  $^{6}$Department of Physics and Astronomy, College of Charleston, 58 Coming St., Charleston, SC 29424 \\
  $^{7}$Astronomy Department, Cornell University, Ithaca, NY 14853 \\
  $^{8}$Gemini Observatory, AURA/Casilla 603, La Serena, Chile \\
  $^{9}$NASA/GSFC $^{10}$Department of Physics and Astronomy, Johns Hopkins University, 3400 North
  Charles St., Baltimore, MD 21218 \\}

\date{Accepted 2010 August 5.  Received 2010 August 4; in original form 2010 July 14}

\pagerange{\pageref{firstpage}--\pageref{lastpage}} \pubyear{2010}

\begin{document}

\label{firstpage}

\maketitle

\begin{abstract}
  We use WIRC, IR images of the Antennae (NGC 4038/4039) together with
  the extensive catalogue of 120 X-ray point sources \citep{zez06} to
  search for counterpart candidates.  Using our proven frame-tie
  technique, we find 38 X-ray sources with IR counterparts, almost
  doubling the number of IR counterparts to X-ray sources first
  identified in \citet{cla07}.  In our photometric analysis, we
  consider the 35 IR counterparts that are confirmed star clusters.
  We show that the clusters with X-ray sources tend to be brighter,
  $K_s\approx$ 16 mag, with $(J-K_s)=1.1$ mag.

  We then use archival {\it HST} images of the Antennae to search for
  optical counterparts to the X-ray point sources.  We employ our
  previous IR-to-X-ray frame-tie as an intermediary to establish a
  precise optical-to-X-ray frame-tie with $<0.6$ arcsec rms positional
  uncertainty.  Due to the high optical source density near the X-ray
  sources, we determine that we cannot reliably identify counterparts.
  Comparing the {\it HST} positions to the 35 identified IR star
  cluster counterparts, we find optical matches for 27 of these
  sources.  Using Bruzual-Charlot spectral evolutionary models, we
  find that most clusters associated with an X-ray source are massive,
  $\sim10^6$ M$_{\sun}$, young, $\sim$10$^6$ yr, with moderate
  metallicities, $Z=0.05$.
\end{abstract}


\section{Introduction}

The numerous X-ray point sources and young, massive star clusters in
the Antennae make this galaxy pair an ideal laboratory for studying
the environments of X-ray binaries (XRBs).  Previously, {\it Chandra}
observations revealed 49 X-ray point sources ranging in luminosity
from $10^{38}-10^{40}$ ergs s$^{-1}$ \citep{zez02}.  A 411 ks total
exposure, consisting of six additional pointings spread over two
years, revealed an additional 71 X-ray sources \citep{zez06} down to a
luminosity of $2\times10^{37}$ ergs s$^{-1}$.  While most are XRBs
with either a black hole or neutron star, those sources with $L_X
>10^{39}$ ergs s$^{-1}$ are more unusual objects classified as
ultraluminous X-ray sources (ULXs). Some theories suggest these ULXs
are intermediate mass black holes with masses from 100 -- 10,000
$M_{\sun}$ \citep[e.g.,][]{fab89,zez99,rob00,mak00}, but there remains
a considerable amount of controversy concerning their origins
\citep[e.g.,][and references therein]{kin01, rob07}.

For this study, we use a distance to the Antennae of 19.3 Mpc (for
$H_{0}=75$ km s$^{-1}$ Mpc$^{-1}$), as determined by the spectroscopic
redshift \citep{zez06}. However, there exists some ambiguity
concerning the distance to this galaxy pair.  While a distance derived
from the Type Ia supernova (SN) 2007sr suggests a distance of 22$\pm$3
Mpc \citep{sch08}, analysis of the RGB colors indicate that the
distance can range from 22 Mpc to as low as 13.8 Mpc
\citep{sav04}. While this lower distance would halve our luminosity
and mass estimates, this factor of a few will not affect any
statistical study of their population. Hence, we do not anticipate
these discrepences in the distance to NGC 4038/4039 as being
problematic.

Compact objects tend to be associated with massive star formation,
which some theories suggest is predominant in young stellar clusters
\citep{lad03}.  In previous work on the Antennae we find a close
association between compact objects and clusters, identifying 15
possible IR counterparts to X-ray sources \citep{cla07}.  Many of
these counterparts reside in the spiral arms and 'bridge' region of
the Antennae -- locations predominant in massive star formation.  The
spiral arms wrap around the northern nucleus and extend to the south
of the southern nucleus, while the bridge region is the dense area
connecting the two galaxies.  Those X-ray sources without counterparts
could be compact objects that escaped their parent cluster or remained
behind after their cluster dissolved.  In \citet{cla07} we suggest a
third possibility, that some X-ray sources do have counterparts, but
these are too faint to see in the IR images.

In this paper, we expand on our initial study by searching for IR
counterparts to the 120 X-ray sources identified in \citet{zez06}.  We
then extend this work to optical wavelengths using {\it HST} images of
the Antennae.  The higher sensitivity of {\it HST} in non-dust
obscured regions allows us to search for additional counterparts to
X-ray sources.  In addition, combining multi-band photometry in the
optical and IR, we can use spectral evolutionary models to measure
cluster properties.  In \S2 we discuss our IR observations and
analysis of optical {\it HST} archival images.  We describe our
photometric analysis of counterpart cluster properties in \S3, and
summarize our results in \S4.

\section{Observations and Data Analysis}

\begin{table*}
 \centering
 \begin{minipage}{123mm}
   \caption{IR Cluster Counterparts to {\it Chandra} X-Ray Sources.
     $\Delta\alpha$ and $\Delta\delta$ are the positional offsets in
     units of arcsec from the {\it Chandra} coordinates to the WIRC
     coordinates of the proposed counterpart.  {\it Chandra} Src ID
     numbers follow the naming convention of \citet{zez06}.  The
     listed RA and Dec {\it Chandra} coordinates have an uncertainty
     of 0.5 arcsec \citep{zez06}.  The near-IR $J$ and $K_s$
     magnitudes were measured using PSF photometry.  The values in
     parenthesis for each magnitude are uncertainties in the final
     listed digit.}
  \begin{tabular}{@{}lcccccc@{}}
    \hline
 {\it Chandra} Src ID & RA &  Dec & $\Delta\alpha$(arcsec) & $\Delta\delta$ (arcsec) & $J$ & $K_s$  \\
  \hline
  Likely \\
  Counterparts \\
  \hline
  7\dotfill & 12:01:49.64 & -18:52:07.10 & 0.45 & 0.66 & 19.93(0.21) & 18.88(0.18) \\
  11\dotfill & 12:01:50.35 & -18:52:15.50 & 0.30 & 0.76 & 16.47(0.02) & 15.89(0.02) \\
  16\dotfill & 12:01:50.51 & -18:52:04.30 & 0.15 & 0.08 & 16.21(0.01) & 15.66(0.02) \\
  22\dotfill & 12:01:51.01 & -18:52:33.00 & 0.45 & 0.55 & 17.56(0.02) & 16.59(0.01) \\
  24\dotfill & 12:01:51.14 & -18:52:31.70 & 0.15 & 0.18 & 17.56(0.02) & 16.59(0.01) \\
  27\dotfill & 12:01:51.32 & -18:52:25.40 & 0.45 & 0.42 & 18.27(0.01) & 17.37(0.08) \\
  38\dotfill & 12:01:52.02 & -18:52:27.50 & 0.60 & 0.30 & 16.82(0.01) & 16.10(0.01) \\
  47\dotfill & 12:01:52.75 & -18:51:30.10 & 0.15 & 0.09 & 18.48(0.04) & 17.78(0.02) \\
  51\dotfill & 12:01:53.00 & -18:52:09.10 & 0.75 & 0.04 & 15.77(0.01) & 15.13(0.06) \\
  52\dotfill & 12:01:53.08 & -18:52:23.80 & 0.60 & 0.52 & 17.16(0.02) & 16.21(0.01) \\
  58\dotfill & 12:01:53.41 & -18:53:07.00 & 0.45 & 0.20 & 15.79(0.18) & 15.51(0.10) \\
  60\dotfill & 12:01:53.42 & -18:53:51.30 & 0.45 & 0.62 & 21.95(1.68) & 17.97(0.09) \\
  83\dotfill & 12:01:54.50 & -18:53:06.70 & 0.00 & 0.57 & 16.71(0.08) & 16.45(0.07) \\
  84\dotfill & 12:01:54.56 & -18:53:03.80 & 0.15 & 0.00 & 15.05(0.03) & 14.27(0.02) \\
  85\dotfill & 12:01:54.62 & -18:52:09.40 & 0.45 & 0.05 & 18.52(0.20) & 16.63(0.01) \\
  86\dotfill & 12:01:54.78 & -18:52:52.00 & 0.15 & 0.11 & 16.76(0.02) & 14.88(0.04) \\
  87\dotfill & 12:01:54.84 & -18:52:14.80 & 0.15 & 0.25 & 16.60(0.03) & 15.92(0.01) \\
  88\dotfill & 12:01:54.98 & -18:53:06.60 & 0.30 & 0.03 & 16.52(0.01) & 14.66(0.01) \\
  91\dotfill & 12:01:55.05 & -18:52:41.00 & 0.90 & 0.50 & 17.81(0.01) & 16.89(0.02) \\
  94\dotfill & 12:01:55.17 & -18:52:47.90 & 0.45 & 0.09 & 17.10(0.07) & 15.71(0.02) \\
  98\dotfill & 12:01:55.54 & -18:52:23.80 & 0.30 & 0.03 & 15.86(0.07) & 15.21(0.04) \\
  99\dotfill & 12:01:55.65 & -18:52:15.20 & 0.60 & 0.13 & 16.30(0.04) & 15.40(0.01) \\
  101\dotfill & 12:01:55.74 & -18:52:42.70 & 0.15 & 0.57 & 16.70(0.02) & 16.10(0.02) \\
  102\dotfill & 12:01:55.74 & -18:52:06.10 & 0.45 & 0.01 & 17.81(0.02) & 17.48(0.02) \\
  107\dotfill & 12:01:56.47 & -18:54:42.00 & 0.15 & 0.06 & --- & --- \\
  117\dotfill & 12:01:58.42 & -18:52:49.60 & 0.30 & 0.80 & 17.64(0.02) & 16.90(0.04) \\
  Possible \\
  Counterparts \\
  \hline
  14\dotfill & 12:01:50.47 & -18:52:21.80 & 0.75 & 1.01 & 15.86(0.03) & 15.12(0.02) \\
  15\dotfill & 12:01:50.47 & -18:52:12.70 & 1.50 & 0.29 & 14.98(0.02) & 14.06(0.01) \\
  36\dotfill & 12:01:51.85 & -18:52:27.80 & 0.75 & 1.25 & 16.77(0.04) & 15.94(0.01) \\
  41\dotfill & 12:01:52.19 & -18:52:20.60 & 1.35 & 0.51 & 17.06(0.04) & 16.18(0.02) \\
  81\dotfill & 12:01:54.35 & -18:52:10.30 & 1.20 & 0.64 & 20.30(0.45) & 16.84(0.03) \\
  89\dotfill & 12:01:54.96 & -18:52:32.50 & 0.15 & 1.19 & 18.11(0.02) & 17.47(0.04) \\
  92\dotfill & 12:01:55.06 & -18:52:16.50 & 0.90 & 1.15 & 17.25(0.03) & 15.77(0.01) \\
  95\dotfill & 12:01:55.37 & -18:52:48.90 & 0.15 & 1.08 & 16.21(0.02) & 15.27(0.04) \\
  119\dotfill & 12:01:59.20 & -18:51:37.30 & 0.60 & 1.01 & 17.78(0.01) & 15.97(0.01) \\
\hline
\end{tabular}
\end{minipage}
\end{table*}

\subsection{Infrared and Optical Imaging}

We base this study on near-infrared (IR) and optical {\it HST} images
of the Antennae.  We report the reduction and analysis of the near-IR
images in \citet{bra05}.  In summary, we acquired 20-minute total
exposures in the $J$ (1.25 $\mu$m) and $K_s$ (2.15 $\mu$m) filters
using the Wide-field InfraRed Camera (WIRC) on the Palomar 5-m
telescope during the night of 2002 March 22.  We obtained the optical
{\it HST} images from the literature \citep{whi99}. This data set was
acquired using the WFPC2 camera and consists of images in the
following filters: F336W($U$), F439W($B$), F555W($V$), and F814W($I$).

In our efforts to understand the environments of the Antennae X-ray
sources, we made frame-ties across three wavelength regimes: {\it
  Chandra} X-ray, WIRC IR and {\it HST} optical images.  Using the IR
as an intermediary between the optical and X-ray frame-tie has many
advantages over direct optical/X-ray matches.  First, the IR
penetrates dust in similar ways to X-rays, facilitating the
identification of counterparts to X-ray sources.  Furthermore, while
bright IR sources may be obscured in the optical, the converse is
generally untrue -- any bright {\it HST} source shows up prominently
in the IR images, enabling an excellent optical/IR frame-tie, and thus
closing the astrometric loop at all wavelengths.  Previous attempts to
match X-ray and optical positions produced many possible counterparts,
but with poor reliability --- as many as 75 percent of the matches are
chance coincidences \citep{zez02}.

\begin{figure*}
\centering
\begin{minipage}[b]{174mm}
\includegraphics[width=97mm]{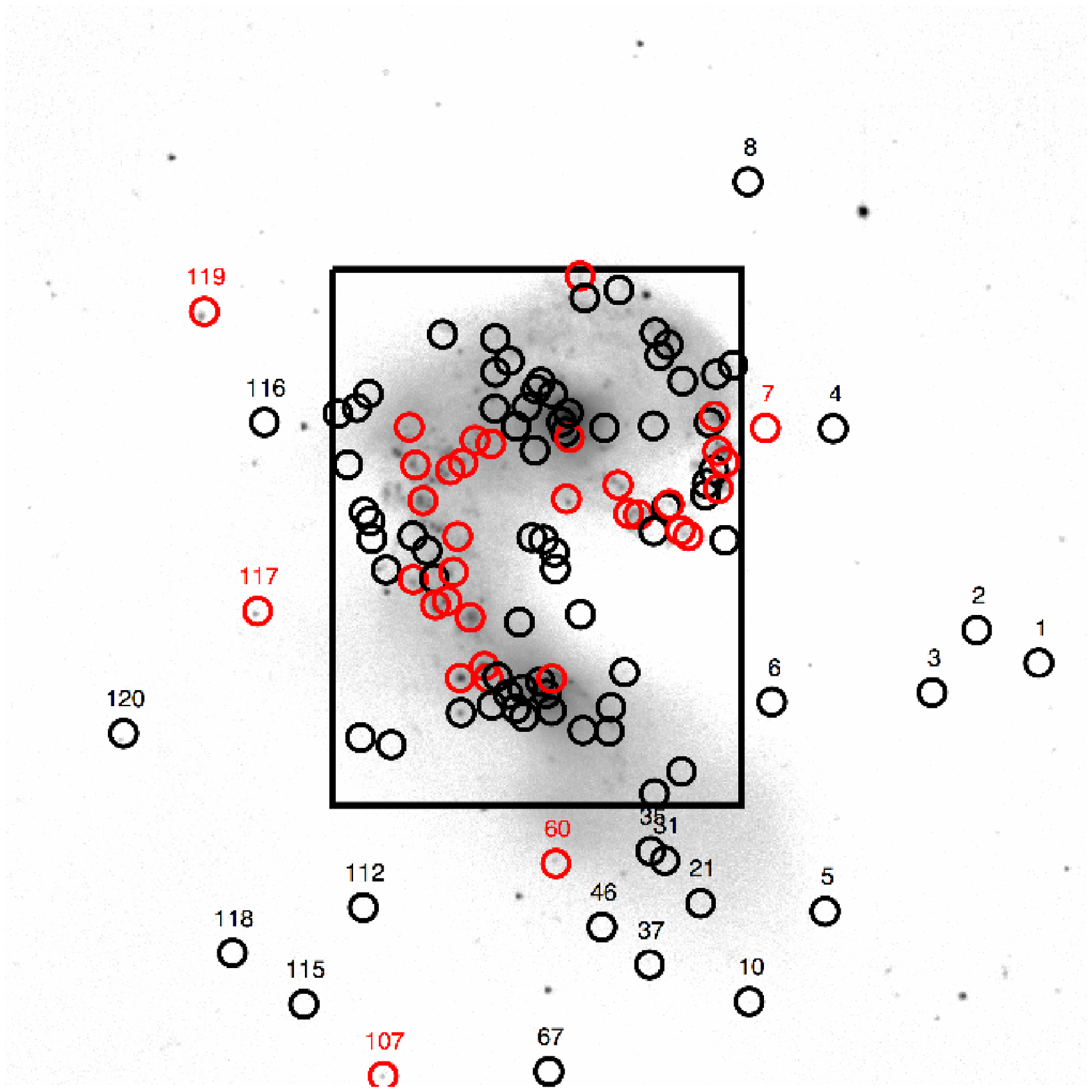}
\includegraphics[width=77mm]{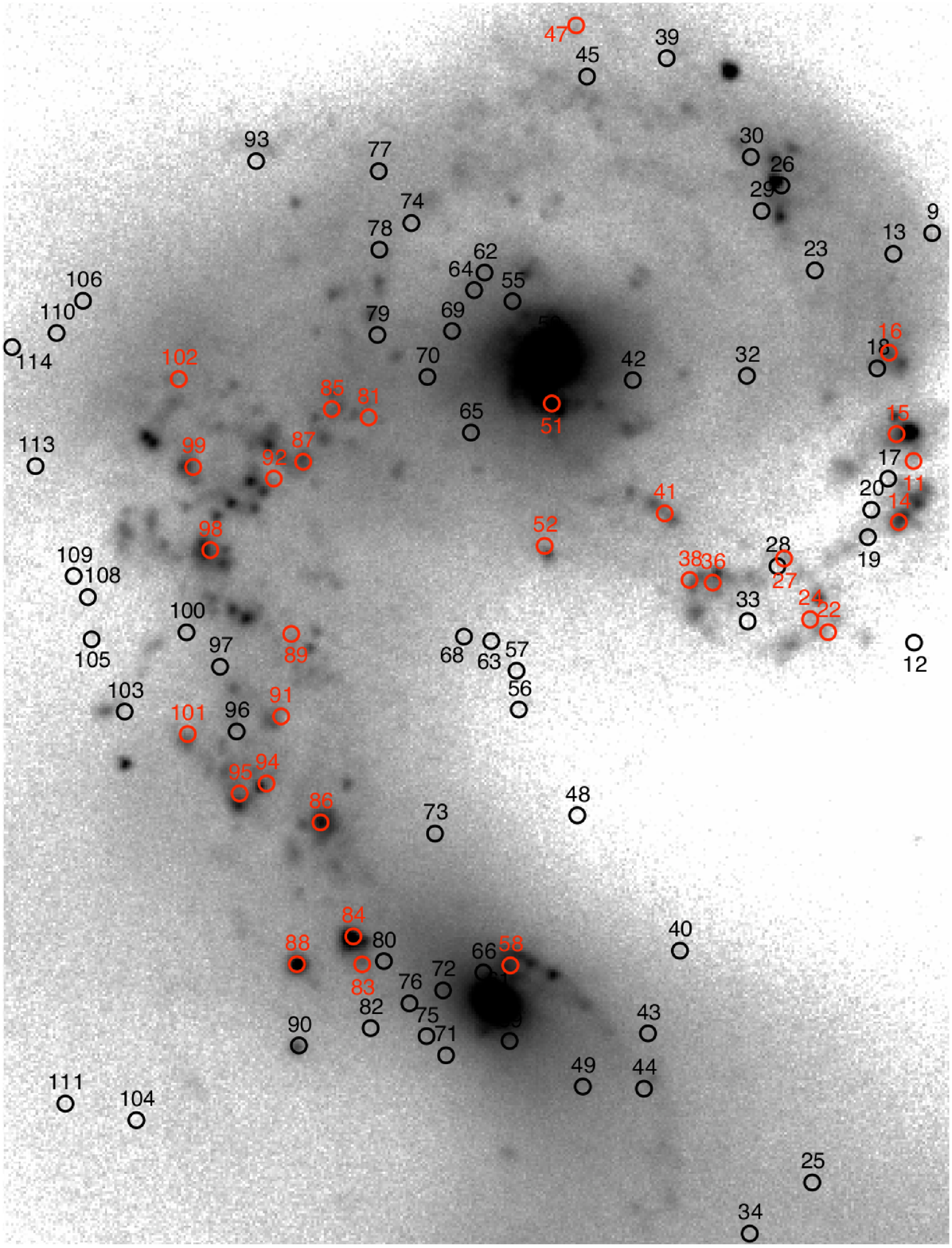}
\caption{X-ray positions from \citet{zez06} overlaid on the WIRC,
  $K_s$ band image of the Antennae.  The right panel is a
  $1.6\times2.0$ arcmin blow-up of the central regions of the galaxy
  pair showing the X-ray source positions in more detail.  Red circles
  are those X-ray sources with IR counterparts, while black circles
  are X-ray sources lacking counterparts.  }
\end{minipage}
\end{figure*}

In the new catalogue of X-ray sources, \citet{zez06} revised the X-ray
source positions and the new positions differ by
$0.2\pm0.4$ arcsec in R.A. and $0.6\pm1.8$ arcsec in Dec
as compared to those coordinates listed in \citet{zez02}.
Considering we are searching for IR counterparts within 1 arcsec of
an X-ray source position (see below), we adjusted our initial
frame-tie \citep{cla07} between the WIRC, IR $K_s$-band image and the
X-ray sources listed in \citet{zez02}, to account for the new, X-ray
coordinates listed in \citet{zez06}.  This modification to the initial
frame-tie yielded the same rms positional uncertainty,
$\sim0.5$ arcsec.

Once our frame-tie between the IR and X-ray was in place, we
extrapolated X-ray positions to optical {\it HST} $I$-band positions.
Tying {\it Chandra} X-ray coordinates directly to {\it HST} positions
is nontrivial due to field crowding in the {\it HST} images.  Instead,
we used our excellent frame-tie between the IR and X-ray as an
intermediary.  Selecting only bright, isolated optical sources, we
made our frame-tie using 12 targets corresponding to sources seen in
the IR.  Using the mapping method describe above, we matched IR pixel
positions to {\it HST} right ascension and declination.  This
frame-tie yielded an rms positional uncertainty of less than
$0.6$ arcsec.  Applying our IR-to-optical astrometric fits to
previously derived $K_s$-band, x,y pixel positions of X-ray sources
\citep{cla07}, we found the {\it HST} positions for all {\it Chandra}
X-ray sources.

\subsection{Identification of Infrared Counterparts to {\it Chandra}
  Sources}
Once our astrometric frame-tie was in place, we found a total of 28
likely and 10 possible IR counterparts to {\it Chandra} X-ray sources
in the Antennae, where likely counterparts are defined to be within a
1.0 arcsec radius (equivalent to twice the positional uncertainty for
our frame-tie) of an X-ray source and possible counterparts are
between 1.0 and 1.5 arcsec (four-to-six times the positional
uncertainty) from an X-ray source.  The 35 sources that are confirmed
star clusters are listed in Table 1 and shown in Fig. 1, with
subimages of each source displayed in Fig. 2.  As can be seen in the
figure, the counterparts are almost exclusively in the spiral arms and
bridge region.  Of the 38 X-ray sources with counterparts, two are the
nuclei \citep[X-50 and X-61 as listed in table 5 of ][]{zez06}, one is
a background quasar \citep[X-90;][]{cla05}, and two lack measured
X-ray luminosities (X-24 and X-52).  The source X-107 is on the edge
of the frame, making sky-subtraction difficult, so we excluded it.
Therefore, in our analysis of cluster properties, we only considered
the 32 IR counterparts that are confirmed star clusters and that have
X-ray sources with measured luminosities.  Furthermore, of the 15 IR
cluster counterparts found in \citet{cla07}, four drop out as
counterparts in this new analysis --- the X-ray sources X-10, X-26,
X-15, and X-22, following the numeration in table 1 of \citet{zez02},
are no longer associated with a star cluster.

\begin{figure*}
\centering
\begin{minipage}[b]{174mm}
\includegraphics[width=87mm]{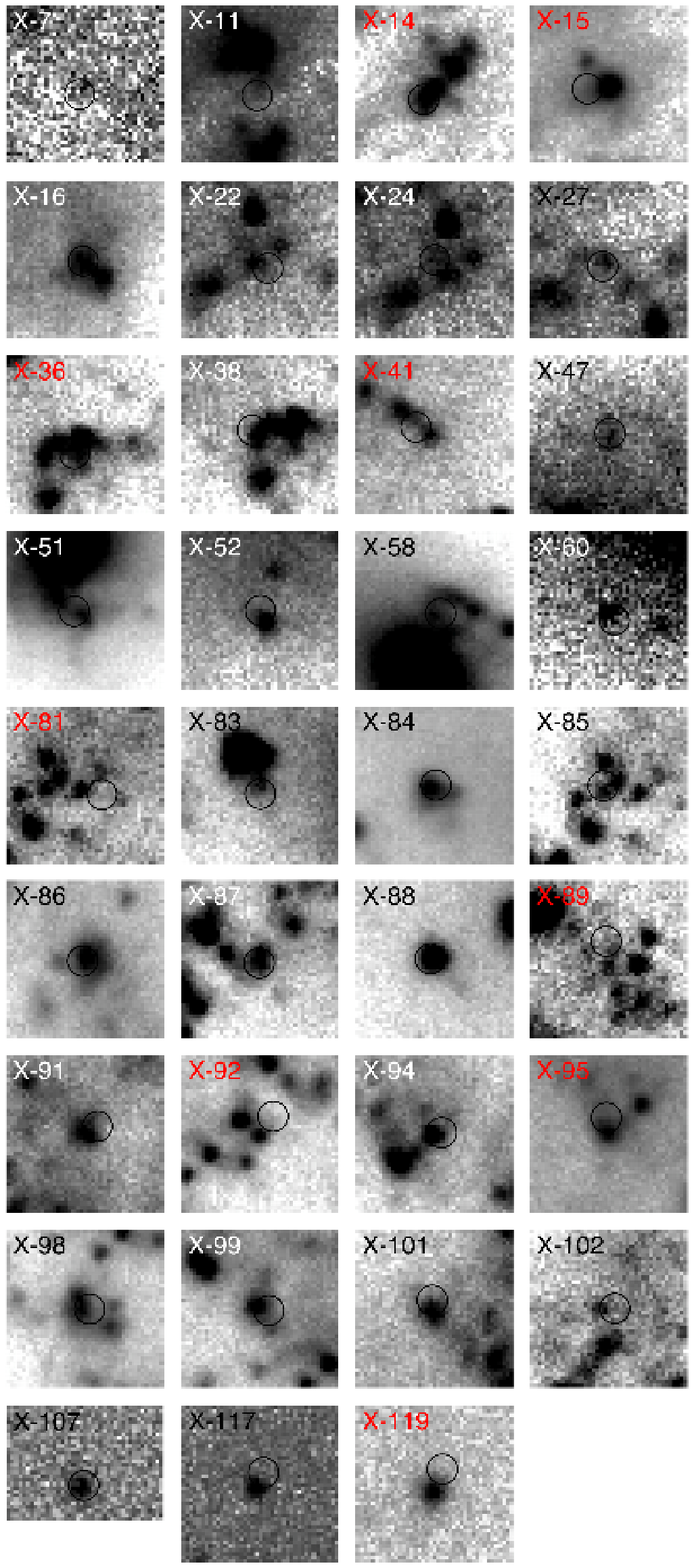}
\includegraphics[width=87mm]{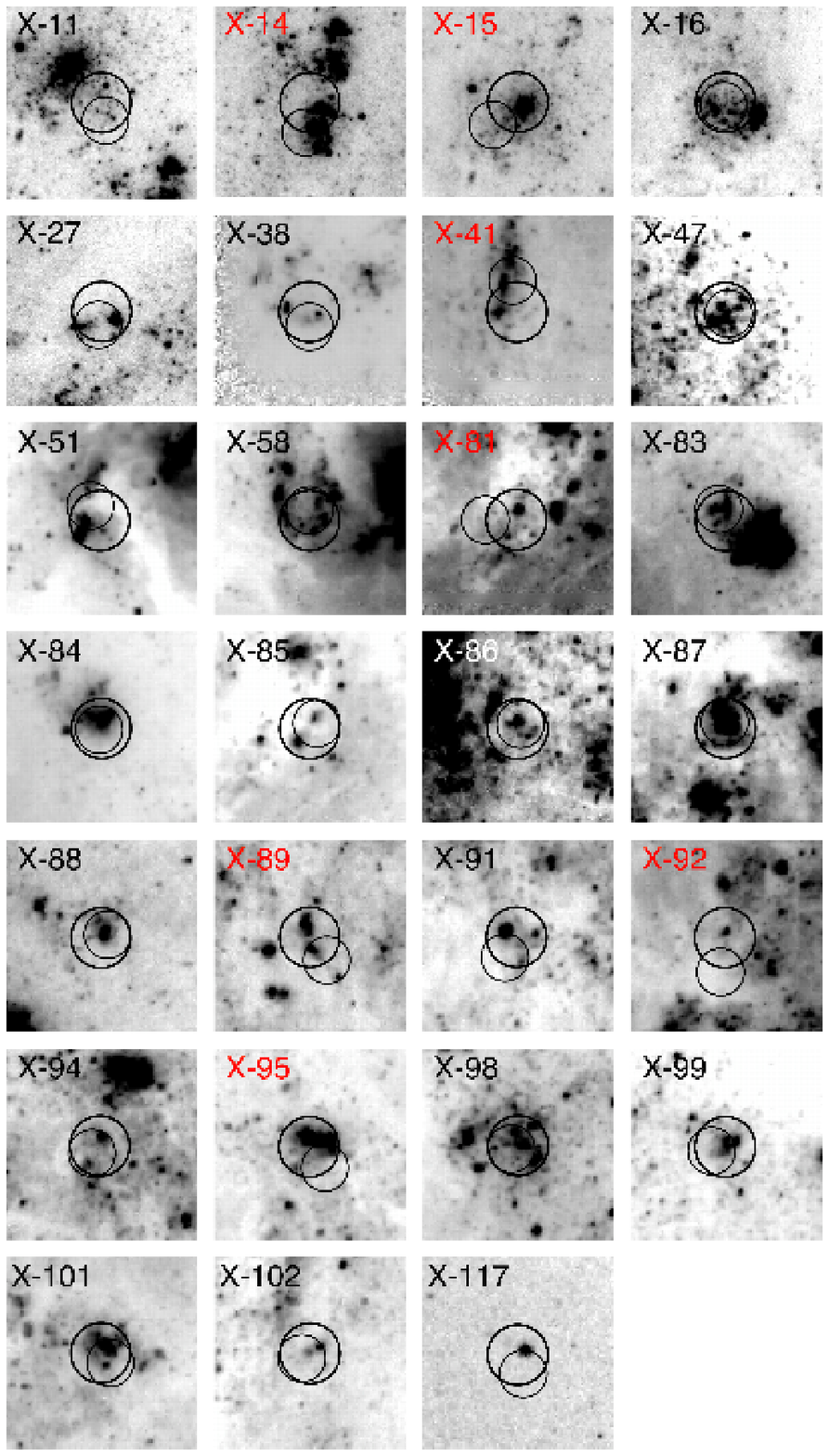}
\caption{(left panel) Subimages from the $K_s$ band image of the
  Antennae for X-ray sources with cluster counterparts.  Each subimage
  is $10\times10$ arcsec, centered on the cluster counterpart and the
  circles are 1 arcsec in radius, centered on the X-ray source
  position.  X-ray sources labelled in red show possible matches to
  cluster counterparts. (right panel) Subimages highlighting optical
  counterparts to X-ray sources with IR cluster counterparts.  The
  small circles are positional error circles with $1.0$ arcsec radii
  centered on X-ray source positions.  The larger circles are centered
  on cluster counterparts and have $1.3$ arcsec radii equivalent to
  the $K_s$-band photometric aperture.  We label those X-ray sources
  with possible counterparts in red.  Each image is
  $6.0\times6.0$ arcsec.  }
\end{minipage}
\end{figure*}

Using the technique developed in \citet{cla07}, which is based on the
IR source density in the field around each X-ray source position, we
attempted to estimate the level of 'contamination' of these
counterpart samples due to chance superpositions of unrelated X-ray
sources and IR clusters.  We expect 8 sources with a $1\sigma$
uncertainty of +0.1/-0.1\footnote{Found using confidence levels for
  small number statistics listed in tables 1 and 2 of \citet{geh86}}
of the 28 likely counterparts to be due to chance superpositions of
unrelated objects, and 10 with a $1\sigma$ uncertainty of
+0.4/-0.3\footnotemark[\value{footnote}] for the 10 possible
counterparts to be chance superpositions.

In Fig. 3 we plot the X-ray luminosity versus the separation in
arcseconds to the associated IR cluster counterpart.  We also divided
the X-ray sources into three luminosity bins: Low Luminosity X-ray
sources (LLX's) had $L_{X}$ $<$ $3\times10^{38}$ ergs s$^{-1}$, High
Luminosity X-ray sources (HLX's) were between $L_{X}$ of
$3\times10^{38}$ergs s$^{-1}$ and $1\times10^{39}$ ergs s$^{-1}$,
while $L_{X}$ $>$ $1\times10^{39}$ ergs s$^{-1}$ were Ultra-Luminous
X-ray Sources (ULX's).  All luminosities were taken from table 5 of
\citet{zez06}.  Notice that there is no trend in separation from the
IR cluster counterpart and X-ray luminosity.  This seems to indicate
no preference as to where XRBs of different luminosity classes form in
star clusters.

\subsection{Identification of Optical Counterparts to {\it Chandra}
  Sources}

The complex field seen in the {\it HST} images of the Antennae makes
finding counterparts to X-ray sources difficult and necessitated a
different method for defining counterparts than that used for the IR
\citep[see above and ][]{cla07}.  Using our precise frame-tie we
defined areas of positional uncertainty around each X-ray source.
Specifically, an inner aperture with a radius of 1 arcsec and an
annular region from 2.0 -- 3.0 arcsec.  We then defined
possible matches as those X-ray sources with two optical sources in
the circular aperture and less than five optical sources in the
annulus, and likely matches as those X-ray sources with one optical
source in the circular aperture and less than five optical sources in
the annulus.  If more than five sources lay in the annular region, we
considered the region to be too complicated for a positive counterpart
identification, regardless of how many sources lay in the inner
aperture.  Using these criteria, we identified seven X-ray sources
with likely matches to a single $I$-band source and one X-ray source,
with possible matches to two $I$-band sources.  We used the {\it HST}
$I$-band image to search for counterparts because this filter covers
the longest wavelength of the available {\it HST} bands and so is
least affected by extinction.

Repeating the procedure discussed in \citet{cla07}, we estimated the
level of source contamination associated with our identified optical
counterparts to X-ray sources.  We expect five with a $1\sigma$
uncertainty of +0.5/-0.3\footnotemark[\value{footnote}] of the seven
likely counterparts to be due to chance superpositions of unrelated
objects, and seven with a $1\sigma$ uncertainty of
+3.8/-5.4\footnotemark[\value{footnote}] for the one possible
counterpart to be chance superpositions.  Clearly these statistics
indicate the majority of our optical counterparts are chance
superpositions and further demonstrate the difficulty of making such
matches in the complex structure of the {\it HST} images of the
Antennae.  Therefore, we did not perform a photometric analysis on
these source as we could not reliably identify the counterparts and so
could not provide any statistically meaningful information on the
X-ray source environments.

Instead, we considered the optical equivalents to the 32 IR cluster
counterparts identified in this work.  These optical matches were
identified using the IR-to-optical astrometric frame-tie to match IR
counterpart positions to {\it HST} positions.  As we discuss below, in
many cases a single IR counterpart split into multiple optical
counterparts, and we labelled these conglomerations as a positive
match.  We found optical counterparts to 27 IR cluster counterparts
and Fig. 2 displays subimages of those counterparts to X-ray sources
seen across all six IR and optical bands.  Fig. 4 is an $I$-band
image of the Antennae showing the positions of all counterpart
candidates to X-ray sources.

\begin{figure}
\includegraphics[width=84mm]{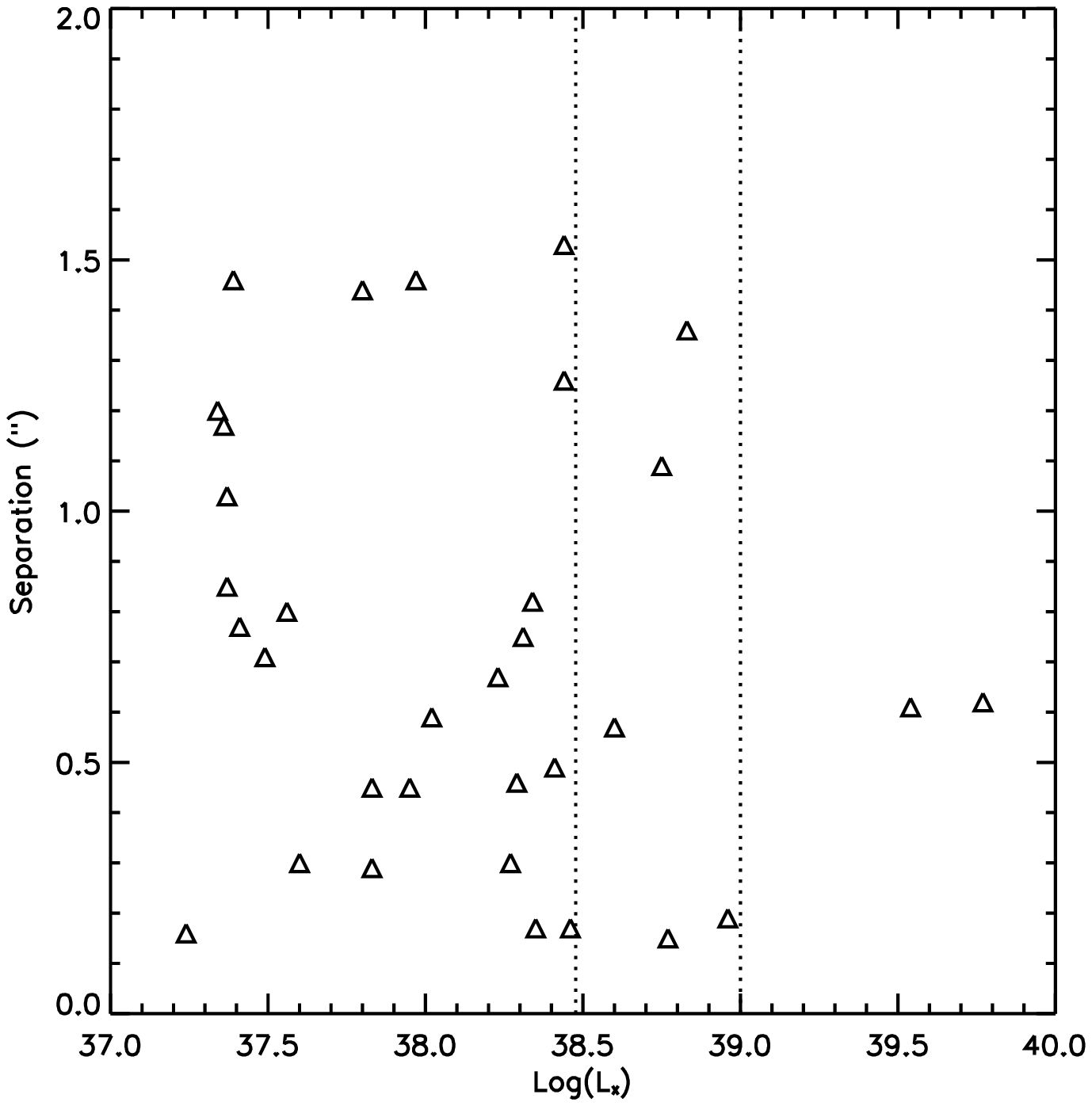}
\caption{Plot of X-ray luminosity versus separation from the IR
  counterpart.  The dashed lines divide the plot into three separate regions: LLX: $L_X$ $<$
  $3\times10^{38}$ ergs s$^{-1}$, HLX: $3\times10^{38}$ ergs s$^{-1}$
  $<$ $L_X$ $<$ $1\times10^{39}$ergs s$^{-1}$, and ULX: $L_X$ $>$
  $1\times10^{39}$ ergs s$^{-1}$.  Notice that there is no trend in X-ray
  luminosity, nor luminosity class, in terms of separation from an IR counterpart.
  }
\end{figure}

\begin{figure}
\includegraphics[width=80mm]{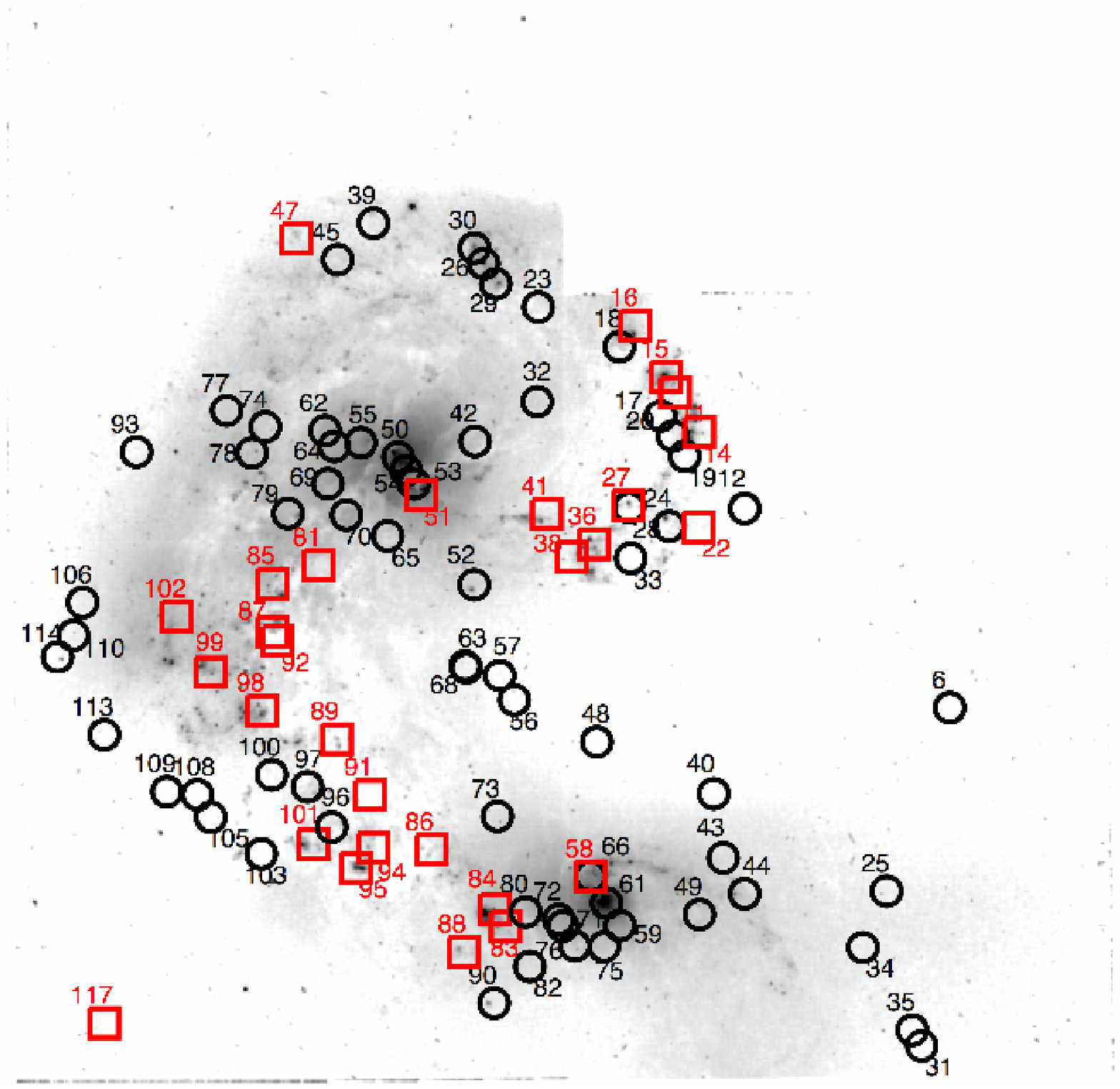} 
\caption{$I$-band Antennae {\it HST} image showing positions of X-ray
  sources from \citet{zez06}.  North is up and east is to the left.  Red squares are
  optical counterparts to X-ray sources with IR cluster counterparts
  identified in this paper.  Black circles are those X-ray sources
  lacking cluster counterparts.
  }
\end{figure}

\section{Results and Discussion}

\subsection{Photometric Properties of the IR Counterparts}

\subsubsection{Colour Magnitude Diagrams}

We made $(J-K_s)$ versus $K_s$ colour magnitude diagrams (CMDs) using
the 226 star clusters with $J$ and $K_s$ magnitudes.  The cluster
magnitudes for 219 sources were first reported in \citet{cla07} and
this study includes photometry for an additional seven clusters.
Fig. 5 expands on the CMD presented in \citet{cla07}, by including a
larger sample of clusters associated with X-ray sources.  As we did
previously \citep{cla07}, we divided the X-ray sources into the three
luminosity classes, LLX, HLX and ULX (see \S2.2 above).  As can be
seen in the plot, the majority of the X-ray sources with IR
counterparts are LLX's, 26, while there are five HLX's and two ULX's.
Interestingly, in the new list of X-ray sources \citep{zez06}, three
that are associated with IR star clusters in both \citet{cla07} and
here change in X-ray luminosity class from those published values in
\citet{zez02}; X-25 goes from HLX to LLX, X-32 from ULX to HLX, and
X-40 from LLX to HLX.  Comparing the old and new sample of IR
counterparts, the new sample is about a half a magnitude fainter (see
Table 2), but shows no difference in colour compared with our previous
study \citep[][table 4]{cla07}.

\begin{table}
 \centering
 \begin{minipage}{70mm}
   \caption{Summary of Potential IR Counterpart Properties.
     $\sigma_{\overline{K}}$ and $\sigma_{\overline{(J-K_s)}}$ are
     uncertainties in each value.}
  \begin{tabular}{@{}lcccccc@{}}
    \hline
Category & $\overline{K_s}$ & $\sigma_{\overline{K}}$ & $\overline{(J-K_s)}$ & $\sigma_{\overline{(J-K_s)}}$ \\
\hline
all clusters & 16.85 & 0.08 & 0.83 & 0.03 \\
X-ray sources & 16.13 & 0.19 & 1.09 & 0.14 \\
LLX & 15.60 & 0.66 & 1.00 & 0.15 \\
HLX & 15.54 & 0.48 & 1.46 & 0.56 \\
ULX & 16.39 & 0.62 & 0.90 & 0.01 \\
\hline
\end{tabular}
\end{minipage}
\end{table}

\subsubsection{Cluster Mass Estimates and $\eta$}

Following the procedure outlined in \citet{cla08}, we computed the
cluster mass relation, $\eta$, defined in that paper, but now included
our extended list of clusters with X-ray counterparts.  As defined in
\citet{cla08}, $\eta$ relates the function of X-ray detections per
mass as a function of cluster mass and can be formalized with the
following equation:

\begin{figure}
\includegraphics[width=80mm]{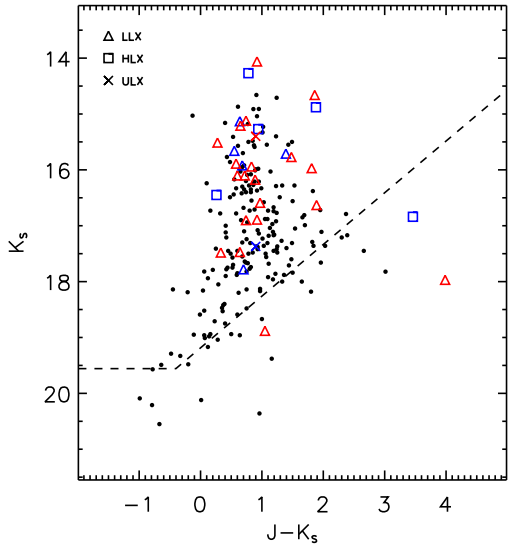} 
\caption{$(J-K_s)$ vs $K_s$ colour-magnitude diagram for all clusters
  in Antennae.  Clusters with X-ray sources are designated by the
  luminosity class of the associated X-ray source.  LLX: $L_X$ $<$
  $3\times10^{38}$ ergs s$^{-1}$, HLX; $3\times10^{38}$ ergs s$^{-1}$
  $<$ $L_X$ $<$ $1\times10^{39}$ergs s$^{-1}$, and ULX: $L_X$ $>$
  $1\times10^{39}$ erg s$^{-1}$.  Blue clusters with X-ray sources
  are from \citet{cla07}, while red clusters with X-ray sources are
  the additional sources added in this work.  The dashed line
    signifies the photometric cut-off we set for statistical purposes.
    }
\end{figure}

\begin{equation}
N_X(M_c) = N_{Cl}(M_c)\cdot\eta(M_c)\cdot M_c
\end{equation}

The quantity, $N_X(M_c)$ is the number of detected X-ray sources with
an IR cluster counterpart, $N_{Cl}(M_c)$ is the number of detected
clusters, and $\eta(M_c)$ is the fraction of X-ray sources per unit
mass, all as a function of cluster mass, $M_c$.

If $\eta(M_c)$ increases or decreases over a range in $M_c$, this
means there could be something peculiar about massive cluster physics
to favour or suppress XRB formation.  In contrast, a constant
$\eta(M_c)$ across all $M_c$ would indicate that more massive clusters
are more likely to have an XRB simply because they have more stars.

We estimated cluster mass using $K_s$-band luminosity, $M_{K_s}$, but
expressed in terms of the observational quantity, flux, $F_{K_s}$.
Given a set of $\eta$ values plotted against cluster flux, a
significant slope could indicate an over-abundance of X-ray sources in
more massive clusters, more than would be expected from simple scaling
arguments.  Fig. 6 shows $\eta(F_{K_s})$ computed for each cluster,
but binned by bins of $F_{K_s}=4.1\times10^6$ DN$^{-1}$.  As can be
seen, there does not appear to be a significant slope and $\eta$ is
consistent with a constant value of $\eta(F_{K_s})=5.5\times10^{-8}$
with an uncertainty of $\sigma_{\overline{\eta}}=7.0\times10^{-9}$ and
is in agreement with the value found in \citet{cla08}.

To estimate the strength of this result, we performed a $\chi^2$ test
between the mean value of $\eta$ and a fitted line to the plotted
values of $\eta$ (Fig. 6) \citep[see][]{cla08}.  We found a
$\Delta\Sigma\chi^2=0.45$. Since any value of $\Delta\Sigma\chi^2$
less than one indicates no significant slope, this suggests there is
no slope in $\eta$.  This differs from the value of
$\Delta\Sigma\chi^2=1.9$ when only using the 15 clusters found with
associations to X-ray sources \citep{cla08} and seems to strengthen
the case that there is only a flat slope in the functional form of
$\eta(M_c)$.

\begin{figure}
\includegraphics[width=80mm]{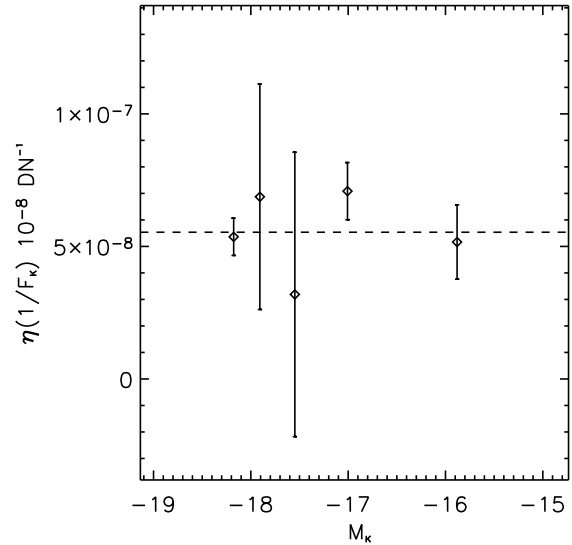} 
\caption{This figure displays $\eta(F_{K_s})$ plotted versus
  $M_{K_s}$.  The bins are $F_{K_s}=4.1\times10^6$ DN$^{-1}$ in
  size.  Error bars are the uncertainties in the mean value of $\eta$
  added in quadrature with the Poisson uncertainty in each bin.  The
  dotted line is the mean of the five $\eta(F_{K_s})$ values.
}
\end{figure}

\subsection{Spectral Evolutionary Models}

During our spectrophotometric analysis of cluster properties, we fit
models to cluster colours across all six IR and optical bands.  We
estimated cluster colours by performing photometry using the same sized
aperture across all bands.  We defined a constant photometric aperture
as $\sim$3$\sigma$ of the $J$-band Gaussian PSF, where the full width
at half maximum (FWHM) is $1.2$ arcsec.  The background annulus had a
radius of $\sim$6--10$\sigma$ of the PSF.  Considering {\it HST}
resolved many of the IR sources into multiple components, this large
photometric aperture encompassed these conglomerations.  Since we
expect that these sources are part of the same, larger structure, it
is appropriate to include photometry of them in spectral evolutionary
models.

A bright, 2MASS star was used to compute $J$ and $K_s$ magnitudes
\citep{bra05}.  We derived {\it HST} magnitudes using zero points
listed in table 28.1 of the {\it HST} Data Handbook \citep{voi97}.
Applying colour transformations defined in \citet{hol95}, we converted
all {\it HST} magnitudes to Johnson $UBVI$ magnitudes.  We expressed
errors in magnitude, $\sigma_m$, as $\sigma_{flux}$ divided by the
mean flux.  In the case of the optical filters, $\sigma_m$ consists of
the additional errors in the zeropoint and colour transformations, all
of which were added in quadrature.

\begin{figure}
\includegraphics[width=80mm]{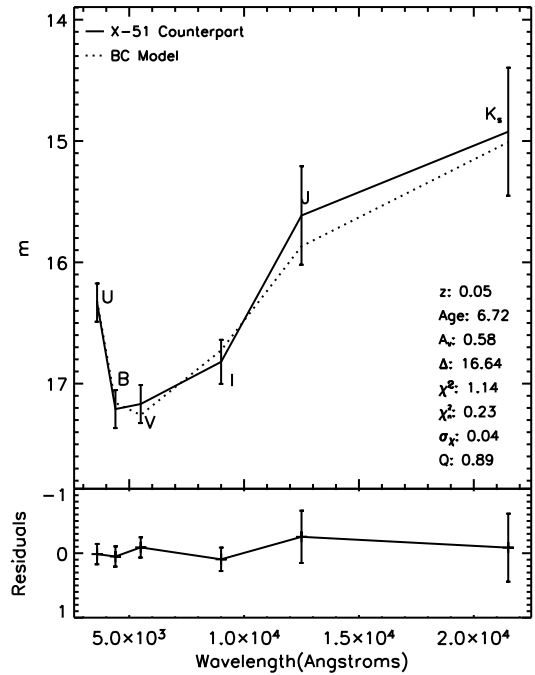} 
\caption{Example of a typical $\chi^2$ fit between a cluster counterpart
  and a BC model.  We include the derived cluster properties and the
  quality of the fit parameters.  Notice the small
  residuals between the model and the data in the lower graph.
  }
\end{figure}

\begin{figure}
\includegraphics[width=86mm]{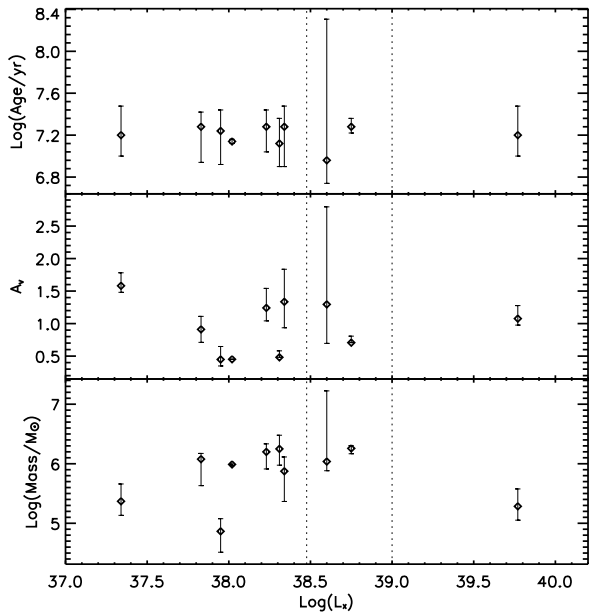}
\caption{Summary of results from $\chi^2$ fitting to BC models.  Error
  bars are ranges in cluster parameters for models within plus
  one of the best-fitting $\chi^2$ sum.  We divide these plots by X-ray
  source luminosity into three separate regions: LLX: $L_X$ $<$
  $3\times10^{38}$ ergs s$^{-1}$, HLX: $3\times10^{38}$ ergs s$^{-1}$
  $<$ $L_X$ $<$ $1\times10^{39}$ergs s$^{-1}$, and ULX: $L_X$ $>$
  $1\times10^{39}$ ergs s$^{-1}$.  Notice that there is no
  obvious trend between cluster properties and the luminosity of the
  associated X-ray source.
  }
\end{figure}

After performing photometry on all X-ray source counterparts seen
across all IR and optical bands, we eliminated six sources suffering
from poor sky subtraction due to crowding, thus resulting in negative
flux measurements.  Therefore, we continued further analysis on 21
counterparts to X-ray sources.  Performing a $\chi^2$ minimization
technique, we fit Bruzual-Charlot \citep[BC;][]{bru03} spectral
evolutionary models to the cluster magnitudes available for each
source to determine mass and age.  We fixed the metallicity to
Z$=0.02$, which is an accepted value for young clusters in the
Antennae \citep{chr08,bas09}. These parameters gave us a more complete
understanding of the XRB environments in the Antennae.  Selecting the
model with the best-fitting $\chi^2$ value, we selected all models that
had a $\chi^2$ within one, $\chi^2+1$, of this value.  The resulting
fits gave us a range in age and reddening for each cluster.  We
estimated reddening by iteratively removing $A_V$ from uncorrected
magnitude cluster SEDs.  By iteratively picking values for $A_V$ in
the range 0.0 -- 3.0 mag and in steps of 0.1 mag, we selected the
$A_V$ that contributed to our best-fitting model.  By allowing the value
of $A_V$ to float, we accounted for measured uncertainties in
reddening.  Each fit is listed in Table 3 and an example fit is shown
in Fig. 7.  We only include those fits with $\chi^2_n<1.6$, which is
the 1$\sigma$ confidence interval in $\chi^2_n$ for the number of
degrees of freedom in our model fits, $n=5$.  $\chi^2_n$ is equal to
$\chi^2/n$.

The shift in magnitude between the model and the data ($\Delta$) gave
us a mass estimate for each IR/optical counterpart.  $\Delta$ contains
information on the distance modulus and mass of the cluster.
Subtracting off the distance modulus to the Antennae, $m_d=31.4$ mag
(for $H_{0}=75$ km s$^{-1}$ Mpc$^{-1}$), we were left with a
difference in luminosity between our cluster and a 1 $M_{\sun}$ cluster
as listed by the BC models.  Converting this difference into a change
in flux gave us the cluster mass in solar masses.  Before computing
this change in flux, we renormalized the luminosity difference,
expressing it in terms of magnitudes instead of colours  We
renormalized the luminosity difference to $M_{K_s}$ for all fits,
where $M_{K_s}$ was computed using PSF, aperture photometry
\citep[see][]{cla07}.

\begin{table*}
 \centering
 \begin{minipage}{118mm}
   \caption{Bruzual-Charlot Model Fits.  {\it Chandra} Src ID numbers
     follow the naming convention of \citet{zez06}.  $\chi^2_n$ is the
     normal $\chi^2$ statistic divided by the degrees of freedom for
     the fit; five for all filters.  Q is a goodness-of-fit measure,
     showing the probability that $\chi^2$ statistic exceeds a
     particular value of $\chi$ by chance.}
  \begin{tabular}{@{}cccccccc@{}}
    \hline
{\it Chandra} Src ID & Log($L_X$) & Log(Age/yr) & $A_V$ & Log(M/$M_{\sun}$) & $\chi^2_n$ & Q \\
\hline
11 & 38.34  & 6.900---7.477 & 0.94---1.84 & 5.37---6.11 & 0.71 & 0.47 \\
27 & 39.77  & 7.000---7.477 & 0.98---1.28 & 5.05---5.58 & 0.62 & 0.54 \\
38 & 38.23  & 7.040---7.440 & 1.04---1.54 & 5.91---6.33 & 0.94 & 0.32 \\
51 & 38.31  & 6.900---7.360 & 0.48---0.58 & 5.98---6.48 & 0.44 & 0.70 \\
83 & 38.60  & 6.740---8.306 & 0.70---2.80 & 5.88---7.22 & 0.18 & 0.93 \\
87 & 37.83  & 6.940---7.420 & 0.71---1.11 & 5.63---6.17 & 0.60 & 0.56 \\
89 & 37.34  & 7.000---7.477 & 1.48---1.78 & 5.13---5.66 & 0.58 & 0.58 \\
95 & 38.75  & 7.220---7.360 & 0.71---0.81 & 6.17---6.30 & 1.47 & 0.12 \\
101 & 38.02 & 7.120---7.160 & 0.45---0.45 & 5.97---5.99 & 1.10 & 0.24 \\
102 & 37.95 & 6.920---7.440 & 0.35---0.65 & 4.51---5.08 & 1.39 & 0.14 \\
\hline
\end{tabular}
\end{minipage}
\end{table*}

We summarize the results from all model fits in Fig. 8.  We found
ages between $\sim9\times10^6$ -- $2\times10^7$ yr.  Masses
ranged from $7\times10^5$ -- $2\times10^6$ M$_{\sun}$, with most
$\sim10^6$ M$_{\sun}$.  The extinction varied from $A_V=0.3$
-- 1.5 mag.  Finally, we did not find a trend between these cluster
properties and the luminosities of their associated X-ray sources.

\section{Conclusions}

In this work, we made astrometric frame-ties between {\it Chandra}
X-ray coordinates \citep{zez06}, WIRC IR pixel positions, and {\it
  HST} optical coordinates to search for star cluster counterparts to
X-ray point sources.  Using the list of 120 X-ray sources in table 3
of \citet{zez06}, we found 38 of these sources are associated with an
IR counterpart, 35 of which are confirmed star clusters.  This
expanded on previous work in which we found IR star cluster
associations to 15 of the 49 X-ray sources listed in table 1 of
\citet{zez02} \citep{cla07}.  Our new sample of IR cluster
counterparts includes 11 of the previously identified 15 counterparts
discussed in \citet{cla07}.  Both this past study, and what we present
here, indicate that most X-ray sources, roughly two thirds, do not
spatially coincide with a cluster.  A comparison between X-ray
luminosity and the separation from the associated cluster, does not
show a trend, thus indicating no relation between XRB type and where
it forms in it's parent cluster.

Extending our frame-tie from the X-ray to the optical using the IR as
an intermediary, we were unable to find reliable optical counterparts
due to the crowded {\it HST} fields. This problem of identifying
counterparts to X-ray sources was also encountered in a similar study,
which focused on optical counterparts to ULXs \citep{pta06}.  Thus,
instead, we found the optical counterparts to 27 of the IR star
clusters associated with an X-ray source.  An examination of the
positions of the cluster counterparts in the Antennae indicates that
the majority are in the spiral arms and bridge region between these
two galaxies.  This seems to clearly indicate that those X-ray sources
with counterparts are tied to star formation in these interacting
galaxies.

Using photometry taken from \citet{bra05}, we made an IR, $JK_s$
photometric study of the 33 IR star cluster counterparts whose X-ray
sources have listed luminosities.  We found most of the IR cluster
counterparts are bright, $\sim$16 mag in $K_s$ and in slightly redder
clusters, $(J-K_s)=1.1$ mag as compared to $(J-K_s)=0.8$ mag for the
general population of clusters.  This confirms our results presented
in \citet{cla07}, in which we performed a similar study with the
smaller sample of IR counterparts.

Following the work in \citet{cla08}, we explored the relationship
between cluster mass and the detected number of X-ray sources.  In
\citet{cla08} we defined a function $\eta$ relating the number of
X-ray detections per mass as a function of cluster mass.  We found
$\eta$ is consistent with a mean value of
$\eta(F_{K_s})=5.5\times10^{-8}$.  Using a $\chi^2$ test, we compared
a fitted slope of the plotted $\eta(F_{K_s})$ values to the mean
$\eta(F_{K_s})$ value.  We found a $\Delta\Sigma\chi^2=0.45$.
Considering we found a $\Delta\Sigma\chi^2=1.9$ in \citet{cla08},
our new study shows a stronger relation between $\eta(F_{K_s})$ and
the mean value, effectively ruling out any inclination towards more
X-ray sources residing in more massive clusters other than through
simple scaling arguments.

Including the 27 star cluster counterparts to X-ray sources seen
across all $UBVIJK_s$ bands, we fit Bruzual-Charlot
\citep[BC;][]{bru03} spectrophotometric models to 10 of these
clusters.  The BC model fits indicate the X-ray-source-associated
clusters are $7\times10^5$ -- $2\times10^6$ M$_{\sun}$ in mass,
$\sim9\times10^6$ -- $2\times10^7$ yr in age, with extinction varying
between $A_V=0.3$ -- 1.5 mag.  These proprieties indicate star
cluster counterparts to X-ray sources in the Antennae tend to be young
and massive, which are consistent with these X-ray sources being
associated with star formation.  In \citet{pta06}, these authors also
found similar mass results for their counterparts to ULXs.

While we can use multiwavelength photometry to describe cluster
properties, there remains some uncertainty in these characteristics
due to errors in magnitude and model limitations.  In future work, we
plan to acquire spectra of Antennae cluster counterparts and refine
estimations of their properties.

\section*{Acknowledgments}

The authors thank the staff of Palomar Observatory for their excellent
assistance in commissioning WIRC and obtaining these data.  WIRC was
made possible by support from the NSF (NSF-AST0328522), the Norris
Foundation, and Cornell University.  Based on observations made with
the NASA/ESA Hubble Space Telescope, obtained from the data archive at
the Space Telescope Science Institute. STScI is operated by the
Association of Universities for Research in Astronomy, Inc. under NASA
contract NAS 5-26555.  DMC is grateful for the many useful discussions
with Michelle Edwards and Valerie Mikles.  SSE and DMC received
support in part by an NSF CAREER award (NSF-9983830) and an NSF grant
(NSF-AST0507547).  We also thank J. Houck for his support of the WIRC
instrument project.

\label{lastpage}

\end{document}